\documentclass[12pt]{article}

\def\beq{\begin{eqnarray}}
\def\eeq{\end{eqnarray}}

\def\L*{{\cal L}_*}
\def\lsim{\mathrel{\rlap{\lower3pt\hbox{\hskip0pt$\sim$}}
     \raise1pt\hbox{$<$}}}         
\def\gsim{\mathrel{\rlap{\lower4pt\hbox{\hskip1pt$\sim$}}
     \raise1pt\hbox{$>$}}}         

\usepackage{amsmath}
\usepackage{amsfonts}

\begin{document}
\begin{titlepage}

\centerline{\Large \bf Gravitational Higgs Mechanism and Massive Gravity}
\medskip

\centerline{\large Zurab Kakushadze}

\bigskip

\centerline{\em 200 Rector Place, Apt 41F, New York, NY 10280}
\centerline{\tt zura@kakushadze.com}
\centerline{(September 25, 2007)}

\bigskip
\medskip

\begin{abstract}
{}In \cite{KL}, in the context of domain wall backgrounds, it was shown that
spontaneous breaking of diffeomorphism invariance results in gravitational Higgs mechanism.
Recently in \cite{thooft} 't Hooft discussed gravitational Higgs mechanism in the context of obtaining
massive gravity directly in four dimensions, and pointed out some subtleties with unitarity. We
attribute these subtleties to breaking time-like diffeomorphisms, and discuss
gravitational Higgs mechanism with all but time-like diffeomorphisms spontaneously broken.
The resulting background is no longer flat but exhibits expansion, which is linear in time.
For space-time dimensions $D \leq 10$ the background
is stable and has no non-unitary propagating modes.
The absence of non-unitary modes is due to the unbroken time-like diffeomorphism invariance.
The physical states correspond to those of a massive graviton. The effective mass squared of the graviton
is positive for $D < 10$, and vanishes for $D = 10$. For $D > 10$ the graviton modes become effectively
tachyonic. The special value of $D=10$, which coincides with the critical dimension of
superstring theory, arises in our setup completely classically.
\end{abstract}
\end{titlepage}

\newpage

\section{Introduction and Summary}

{}Unbroken gauge symmetries are associated with massless gauge particles. Photon, to a very
high precision, is one such gauge particle. Some gauge symmetries ({\em e.g.}, electroweak),
on the other hand, are spontaneously broken, and (some of) the corresponding gauge particles
acquire masses via Higgs mechanism. A scalar particle is ``eaten" by a massless gauge boson,
which produces a massive vector boson.

{}General coordinate reparametrization invariance has its own massless particle, a graviton.
Spontaneous breaking of diffeomorphism invariance can then also be expected to be associated
with gravitational Higgs mechanism, where the graviton, or some of its components if such
breaking is incomplete, would acquire mass.

{}Gravitational Higgs Mechanism was discussed in detail in \cite{KL} in the context of domain
wall backgrounds. In such backgrounds diffeomorphisms in the direction transverse to the domain wall
are spontaneously broken by a scalar field. This then results in gravitational Higgs mechanism. Thus,
the $D$-dimensional theory has $D(D-3)/2$ graviton modes, plus one scalar mode. The scalar fluctuations
can be gauged away by the diffeomorphism in the transverse direction. Graviphotons can be gauged away
using the remaining $(D-1)$ diffeomorphisms. The remaining graviscalar component cannot be gauged away
but has no normalizable (neither plain-wave nor quadratically normalizable) modes. If the domain wall
interpolates between two AdS vacua (finite volume in the transverse direction), we have one quadratically
normalizable massless $(D-1)$-dimensional graviton mode, and a continuum of plain-wave normalizable
massive $(D-1)$-dimensional graviton
modes\footnote{Finite volume solutions with explicitly broken diffeomorphism
invariance were originally discussed in \cite{RS}.}. If the domain wall
interpolates between an AdS vacuum and a Minkowski vacuum (infinite volume in the
transverse direction), we have a continuum of plain-wave normalizable
$(D-1)$-dimensional graviton modes, including the massless one.

{}One of the key points of \cite{KL} is that in backgrounds with {\em spontaneously} (as opposed to
explicitly) broken diffeomorphisms extra modes can be gauged away using these diffeomorphisms as the
equations of motion are invariant under the full diffeomorphism invariance of the theory. Subsequently,
gravitational Higgs mechanism was discussed in various contexts, see, {\em e.g.}, \cite{Por, Ch, Kir}. 
For earlier works, see, {\em e.g.}, \cite{GT, Siegel}. For
a recent review of massive gravity in the context of infinite volume extra dimensions, see, {\em e.g.},
\cite{Gab} and references therein. For a recent review of spontaneous breaking of diffeomorphism symmetry
in the context of Lorentz violating Chern-Simons modification of gravity, see \cite{Jackiw} and references
therein.

{}Recently, 't Hooft discussed gravitational Higgs mechanism in the context of obtaining massive
gravity directly in four dimensions \cite{thooft}\footnote{I would like to thank Olindo Corradini for pointing
out 't Hooft's paper.}. One of the motivations for 't Hooft's work,
and a very compelling one, is actually QCD. If QCD is to be described by string theory, all known
consistent versions of which contain massless gravity, then the graviton should presumably somehow acquire
mass. Gravitational Higgs mechanism is one way of approaching this problem.

{}Thus, 't Hooft considered a four-dimensional background where diffeomorphisms are broken spontaneously by
four scalar fields whose VEVs are proportional to the four space-time coordinates. This is not
a static background as one of the four scalars is time-dependent. Einstein's equations then have a flat
solution if a negative cosmological constant term is introduced. Linearized gravity in this background is
massive, but one non-unitary mode (the trace of the spatial part of the graviton) is also propagating. As we
discuss in the following, the reason why is that the massless graviton has two propagating degrees of freedom,
while the massive one has five. There are four scalars in this setup, and only three can be ``eaten" in the
gravitational Higgs mechanism. There is therefore an extra non-unitary degree of freedom, which does
not decouple. The reason for this non-unitarity can be traced to the fact that one of the four scalars, the one
that breaks time-like diffeomorphisms, is
(effectively) time-like. In \cite{thooft} two ways of removing this non-unitarity were discussed.

{}The above count of propagating degrees of freedom suggests the following approach. Since, in four dimensions, the
massless graviton has two propagating degrees of freedom, and the massive graviton has five propagating degrees
of freedom, three scalars should suffice for gravitational Higgs mechanism. In this note we discuss precisely such
a setup, where in $D$ dimensions $(D-1)$ scalars spontaneously break diffeomorphism invariance in all of the {\em spatial}
directions\footnote{One other difference from 't Hooft's case is that the cosmological constant here is vanishing.}.
The resulting background is not a Minkowski space but a conformally flat expanding
background\footnote{So, just
as in 't Hooft's case this background is not static, but here it is the metric that is time dependent, while in 't
Hooft's case it was one of the scalars.}. We analyze small fluctuations in this background and show that the only
propagating degrees of freedom are indeed $(D+1)(D-2)/2$ components of a massive graviton in $D$ dimensions. So,
gravitational Higgs mechanism works exactly as expected without any non-unitary propagating degrees of freedom.

{}An interesting feature of our model is that for $D > 10$ the effective mass squared of the graviton becomes negative,
{\em i.e.}, the graviton modes become effectively tachyonic. The effective mass squared of the graviton is positive for
$D < 10$, and it {\em vanishes} for $D = 10$. The special value of $D=10$, which coincides with the critical dimension of
superstring theories, arises in our setup completely classically. At present it is unclear if there is a deeper connection
here, which would be interesting to understand.

{}Since our background is not static, it is not clear if it is directly applicable to the aforementioned QCD related
motivation, albeit the connection to $D=10$ is intriguing even in this context.
In this regard it would be interesting to see if one can
construct static, and perhaps even flat, backgrounds with spontaneously
broken (spatial) diffeomorphisms where we expect to have massive gravity via gravitational Higgs mechanism. However, this is
beyond the scope of this note, whose purpose is simply to illustrate how gravitational Higgs mechanism works in this
context. On the other hand, our findings could perhaps have implications for the cosmological constant problem. In particular,
our background exhibits linear (as opposed to exponential as in the positive cosmological constant case) expansion while the
cosmological constant is actually vanishing. The corresponding length scale is set by the spontaneous symmetry breaking. This
might be one approach to avoiding fine tuning. In particular, it would be interesting to see if this can be useful in the
context of the accelerating universe \cite{Acc}.

\section{Spontaneous Symmetry Breaking}

{}Consider $d$ real scalar fields $\phi^a$ ($a = 1,\dots,d$) coupled to gravity
with the following action:
\begin{equation}
 S=M_P^{D-2}\int d^Dx \sqrt{-G}\left[ R - \nabla^M \phi^a \nabla_M \phi_a \right]~,
 \label{actionphi}
\end{equation}
where $M_P$ is the $D$-dimensional (reduced) Planck scale. The space-time coordinates $x^M$,
$M=0,\dots,D-1$ have metric with signature $(-,+,\dots,+)$. The indices $a,b,\dots$ are raised
and lowered with Euclidean metric $\delta^{ab}$, $\delta_{ab}$, so the scalar sector possesses
$SO(d)$ global symmetry. In the following we will be interested in cases where $d = D - 1$.

{}The equations of motion read:
\begin{eqnarray}
 \label{phi1}
 && \nabla^2 \phi^a = 0~,\\
 \label{einstein1}
 && R_{MN} - {1\over 2}G_{MN} R
 = \nabla_M\phi^a \nabla_N\phi_a
 -{1\over 2}G_{MN}\nabla^S \phi^a \nabla_S \phi_a ~.
\end{eqnarray}
In the following we will be interested in solutions that break (part of) the $D$-dimensional
diffeomorphisms spontaneously. Thus, solutions with
\begin{equation}
\label{linphi}
 \phi^a = m~{\delta^a}_M~x^M~,
\end{equation}
where $m$ is some constant, spontaneously
break diffeomorphisms in the spatial directions while at the same time preserving the $SO(d)$ global
symmetry.

{}It is not difficult to check that (\ref{linphi}) indeed gives a solution to (\ref{phi1})
if the metric has the following conformally flat form ($\eta_{MN}$ is the flat $D$-dimensional Minkowski metric):
\begin{equation}\label{warpedy}
 ds^2=\exp(2A)\eta_{MN}dx^M dx^N~,
\end{equation}
where the warp factor $A$ is independent of the spatial coordinates $x^i$, $i=1,\dots,D-1$, and only depends on the
time coordinate $\tau\equiv x^0$. With this Ansatz we have the following equations of motion
for $A$ (prime denotes derivative w.r.t. $\tau$):
\begin{eqnarray}
 \label{phi'A'1}
 &&(D-2)(A^\prime)^2 = m^2~,\\
 \label{A''1}
 &&A^{\prime\prime} = 0~.
\end{eqnarray}
Here the first equation follows from the $(00)$ component
of (\ref{einstein1}), and the second equation follows from a linear combination of (\ref{phi'A'1}) and the $(ij)$
component of (\ref{einstein1}). Our background is therefore given by (\ref{linphi}) and
\begin{equation}
 \label{sol-A}
 A(\tau) = {m\over\sqrt{D-2}}~(\tau - \tau_0)~,
\end{equation}
where $\tau_0$ is an integration constant. Note that this expanding solution is {\em not} the same as that in the
positive cosmological constant case. Thus, instead of the coordinates $\tau, x^i$ let us switch to the coordinates $t, x^i$,
where the time coordinate $t$ in the metric is not warped:
\begin{equation}
 ds^2 = -dt^2 + {m^2\over{D-2}}~(t-t_0)^2~dx^i dx_i~,
\end{equation}
where $t_0$ is an integration constant. In the positive cosmological constant case the corresponding expansion factor is
actually exponential. Moreover, in the positive cosmological constant case the scalar curvature is constant, while in our case
it is time-dependent:
\begin{equation}
 R = (D-1)(D-2)(t-t_0)^{-2}~.
\end{equation}
In the following, in studying the propagating modes in this background, we will assume that we are far enough into the future
away from the ``crunch" point $t = t_0$.

\section{Propagating Modes}

{}In this section we discuss the physical modes propagating in the background of the previous section.
Thus, let us consider small fluctuations around the metric (\ref{warpedy})
\begin{equation}\label{fluctu}
 G_{MN}=\exp(2A)\left[\eta_{MN}+{\widetilde h}_{MN}\right]~,
\end{equation}
where for convenience reasons we have chosen to work with
${\widetilde h}_{MN}$
instead of metric fluctuations $h_{MN}=\exp(2A){\widetilde h}_{MN}$.
Also, let $\varphi^a$ be the fluctuations of the scalar fields around the
background (\ref{linphi}).

{}In terms of ${\widetilde h}_{MN}$ the full
$D$-dimensional diffeomorphisms (corresponding to $x^M\rightarrow x^M-\xi^M$)
\begin{equation}
 \delta h_{MN}=\nabla_M\xi_N+\nabla_N\xi_M
\end{equation}
are given by the following gauge
transformations (here we use $\xi_M\equiv \exp(2A){\widetilde \xi}_M$, and the indices on
tilded quantities are raised and lowered with the Minkowski metric $\eta^{MN}$, $\eta_{MN}$):
\begin{equation}\label{gauge}
 \delta{\widetilde h}_{MN}=\partial_M {\widetilde\xi}_N+
 \partial_N{\widetilde\xi}_M + 2A^\prime\eta_{MN} n^S {\widetilde \xi}_S~,
\end{equation}
where we have introduced a unit vector $n_M\equiv(1,0,\dots,0)$,
$n^M\equiv(-1,0,\dots,0)$. As to the scalar fields $\varphi^a$,
we have:
\begin{equation}\label{diffphi}
 \delta\varphi^a =\nabla_M \phi^a {\widetilde \xi}^M = m~{\delta^a}_i ~{\widetilde \xi}^i~.
\end{equation}
Since our solution does not break diffeomorphisms explicitly
but spontaneously, the linearized equations of motion are invariant under the full
$D$-dimensional diffeomorphisms.

{}Let us count the number of physical degrees of freedom. Thus, we have Einstein-Hilbert gravity
with $D(D-3)/2$ propagating degrees of freedom, plus $d = D-1$ scalars. The total number of
propagating degrees of freedom is $(D+1)(D-2)/2$, which is the number of degrees of freedom in
massive $D$-dimensional gravity. In fact, as we will see in the following, spontaneous breaking of
the diffeomorphism symmetry indeed results in physical degrees of freedom corresponding to
massive gravity.

{}Let us now see this in more detail. In the following we will keep only first order terms in ${\widetilde h}_{MN}$
and $\varphi^a$ in the equations of motion. Next, the linearized equation of motion (\ref{phi1}) and
(\ref{einstein1}) read:
\begin{eqnarray}
 &&\partial^M\partial_M\varphi^a + (D-2)A^\prime n^S\partial_S\varphi^a +\nonumber\\
 \label{EOMphi}
 &&{m \over 2} ~{\delta^a}_S \left[  \partial^S {\widetilde h} - 2\partial_N {\widetilde h}^{SN}
 -2(D-2)A^\prime n_N{\widetilde h}^{SN}\right] = 0~,\\
 &&\left\{\partial_S\partial^S {\widetilde h}_{MN} +\partial_M\partial_N
 {\widetilde h}-\partial_M \partial^S {\widetilde h}_{SN}-
 \partial_N \partial^S {\widetilde h}_{SM}-\eta_{MN}
 \left[\partial_S\partial^S {\widetilde h}-\partial^S\partial^R
 {\widetilde h}_{SR}\right]\right\}+\nonumber\\
 &&(D-2)A^\prime\left\{\left[\partial_S {\widetilde h}_{MN} -
 \partial_M {\widetilde h}_{NS}-\partial_N{\widetilde h}_{MS}\right] n^S
 +\eta_{MN}\left[2\partial^R {\widetilde h}_{RS} - \partial_S
 {\widetilde h}\right] n^S\right\} + \nonumber\\
 \label{EOMh0}
 &&(D-2)(D-3)(A^\prime)^2\left[{\widetilde h}_{MN} +
 \eta_{MN}{\widetilde h}_{SR}n^S n^R \right] = \nonumber\\
 &&m^2 \left\{(D-1){\widetilde h}_{MN} - \eta_{MN}\left[{\widetilde h} +
 {\widetilde h}_{SR}n^S n^R \right]\right\} + \nonumber\\
 &&2m\left[\eta_{MN}{\delta_a}^S\partial_S\varphi^a -
 {\delta^a}_M\partial_N\varphi_a - {\delta^a}_N\partial_M\varphi_a\right]~,
\end{eqnarray}
where ${\widetilde h}\equiv{\widetilde h}^M_M$.
These equations are indeed
invariant under the full diffeomorphism transformations (\ref{gauge}) and (\ref{diffphi}).

{}We can now use the $(D-1)$ diffeomorphisms given by (\ref{diffphi}) to gauge
away the scalar degrees of freedom $\varphi^a$:
\begin{eqnarray}\label{zerophi}
 &&\varphi^a = 0~,\\
 \label{extrah}
 &&{\delta^a}_S \left[\partial^S {\widetilde h} - 2\partial_N {\widetilde h}^{SN}
 -2(D-2)A^\prime n_N{\widetilde h}^{SN}\right] = 0~,\\
 &&\left\{\partial_S\partial^S {\widetilde h}_{MN} +\partial_M\partial_N
 {\widetilde h}-\partial_M \partial^S {\widetilde h}_{SN}-
 \partial_N \partial^S {\widetilde h}_{SM}-\eta_{MN}
 \left[\partial_S\partial^S {\widetilde h}-\partial^S\partial^R
 {\widetilde h}_{SR}\right]\right\}+\nonumber\\
 &&(D-2)A^\prime\left\{\left[\partial_S {\widetilde h}_{MN} -
 \partial_M {\widetilde h}_{NS}-\partial_N{\widetilde h}_{MS}\right] n^S
 +\eta_{MN}\left[2\partial^R {\widetilde h}_{RS} - \partial_S
 {\widetilde h}\right] n^S\right\} + \nonumber\\
 \label{EOMh}
 &&(D-2)(D-3)(A^\prime)^2\left[{\widetilde h}_{MN} +
 \eta_{MN}{\widetilde h}_{SR}n^S n^R \right] = \nonumber\\
 &&m^2 \left\{(D-1){\widetilde h}_{MN} - \eta_{MN}\left[{\widetilde h} +
 {\widetilde h}_{SR}n^S n^R \right]\right\}
\end{eqnarray}
After this gauge fixing, we have one diffeomorphism remaining, that given by $n^S{\widetilde \xi}_S$.

{}Note that our background is time dependent, and there are terms involving $A^\prime$ in
the above equations of motion. Luckily, however, in our background $A^\prime$ is a constant,
so (\ref{EOMh}) is just a second order
equation with constant coefficients. The corresponding propagator is therefore perfectly well behaved in the sense that it
should not
have any unexpected singularities. However, we must still make sure that there are no non-unitary propagating degrees of freedom,
and that there are no tachyonic modes.

{}Let us begin by simplifying the above equation by utilizing (\ref{phi'A'1}) and combining the mass terms:
\begin{eqnarray}
 &&\left\{\partial_S\partial^S {\widetilde h}_{MN} +\partial_M\partial_N
 {\widetilde h}-\partial_M \partial^S {\widetilde h}_{SN}-
 \partial_N \partial^S {\widetilde h}_{SM}-\eta_{MN}
 \left[\partial_S\partial^S {\widetilde h}-\partial^S\partial^R
 {\widetilde h}_{SR}\right]\right\}+\nonumber\\
 &&(D-2)A^\prime\left\{\left[\partial_S {\widetilde h}_{MN} -
 \partial_M {\widetilde h}_{NS}-\partial_N{\widetilde h}_{MS}\right] n^S
 +\eta_{MN}\left[2\partial^R {\widetilde h}_{RS} - \partial_S
 {\widetilde h}\right] n^S\right\}=\nonumber\\
 \label{EOMh1}
 &&m^2 \left\{2{\widetilde h}_{MN} - \eta_{MN}\left[{\widetilde h} + (D-2)
 {\widetilde h}_{SR}n^S n^R \right]\right\}~.
\end{eqnarray}
Next, let
\begin{equation}
 Q^S \equiv \partial_N {\widetilde h}^{SN} -{1\over 2} \partial^S {\widetilde h}
 +(D-2)A^\prime n_N{\widetilde h}^{SN}~.
\end{equation}
According to (\ref{extrah}), the spatial components of this vector vanish. Note that under the
full diffeomorphisms we have the following transformation property:
\begin{equation}\label{deltaQ}
 \delta Q_S = \partial^N\partial_N \xi_S + (D-2) A^\prime n^R\partial_R\xi_S + 2m^2 n_S n^R\xi_R~.
\end{equation}
This implies that we can use the remaining (after the gauge fixing (\ref{zerophi}))
time-like diffeomorphism $n^S\xi_S$ to set $n^S Q_S$ to zero. We then have:
\begin{equation}\label{Q}
 Q_S = 0~.
\end{equation}
Plugging this into (\ref{EOMh1}), we obtain the following diagonalized equation:
\begin{eqnarray}
 \label{EOMh3}
 \partial_S\partial^S {\widehat h}_{MN} +
 (D-2)A^\prime n^S \partial_S {\widehat h}_{MN} = 2 m^2 {\widehat h}_{MN} ~,
\end{eqnarray}
where
\begin{equation}
 {\widehat h}_{MN} \equiv {\widetilde h}_{MN} - {1\over 2} \eta_{MN} {\widetilde h}~.
\end{equation}
This then implies that
\begin{eqnarray}
 \label{EOMh4}
 \partial_S\partial^S {\widetilde h}_{MN} +
 (D-2)A^\prime n^S \partial_S {\widetilde h}_{MN} = 2 m^2 {\widetilde h}_{MN} ~,
\end{eqnarray}
So, at first it might seem that we have $D(D+1)/2 - D = D(D-1)/2$ propagating degrees of freedom, $D(D+1)/2$
components of ${\widetilde h}_{MN}$ less $D$ conditions (\ref{Q}). However, the number of propagating degrees of
freedom is actually one fewer, {\em i.e.}, $(D+1)(D-2)/2$.

{}To see this, recall the transformation property for $n^SQ_S$ from (\ref{deltaQ}):
\begin{equation}\label{deltaQ1}
 \delta (n^S Q_S) = \partial^N\partial_N \Omega + (D-2) A^\prime n^R\partial_R\Omega - 2m^2 \Omega~.
\end{equation}
What this means is that as long as $\Omega \equiv n^S\xi_S$ satisfies the following equation
\begin{equation}
 \partial^N\partial_N \Omega + (D-2) A^\prime n^R\partial_R\Omega = 2m^2 \Omega~,
\end{equation}
the conditions (\ref{Q}) are unaffected. We therefore have {\em residual gauge invariance} in our system. And this
residual gauge invariance corresponds to time-like diffeomorphisms $\Omega$ satisfying the same equation of motion
(\ref{EOMh4}) as the graviton components. In particular, we can use this residual gauge invariance to remove the trace
${\widetilde h}$:
\begin{equation}
 \delta {\widetilde h} = 2\Omega^\prime + 2DA^\prime \Omega~.
\end{equation}
So, we now have:
\begin{eqnarray}
 &&{\widetilde h} = 0~,\\
 &&\partial^R {\widetilde h}_{SR} + (D-2)A^\prime n^R{\widetilde h}_{SR} = 0~.
\end{eqnarray}
We therefore indeed have only $D(D-1)/2 - 1 = (D+1)(D-2)/2$ propagating degrees of freedom, which is what we expect for
a massive graviton in $D$ dimensions.

{}Let us now see what the mass of the graviton is. Let
\begin{equation}
 {\widetilde h}_{MN} = {\rm exp}\left[ -{1\over 2}(D-2)A \right] H_{MN}~.
\end{equation}
In terms of $H_{MN}$ the equations of motion read ($H\equiv H^M_M$):
\begin{eqnarray}
 &&H = 0~,\\
 &&\partial^R H_{SR} + {1\over 2}(D-2)A^\prime n^R H_{SR} = 0~,\\
 \label{EOMh6}
 &&\partial_S\partial^S H_{MN}  = M_H^2 H_{MN}~,
\end{eqnarray}
where
\begin{equation}
 M_H^2 \equiv {{10 - D}\over 4} ~m^2
\end{equation}
is the effective mass squared of the graviton\footnote{We refer to the mass squared appearing in the Klein-Gordon equation
as effective mass squared because our background is actually not static.}.

{}An interesting feature of our solution
is that the graviton modes are massive for $D < 10$, while for $D > 10$ they are effectively tachyonic and (at least
perturbatively) the background becomes unstable. For $D = 10$ the gravity is actually effectively
{\em massless}! Our considerations have been of
purely classical nature, yet we singled out the dimension $D=10$ of critical superstring theory, which is a quantum
requirement. It would be interesting to understand if there is indeed a deeper connection here.

{}Before we end this section let us make the following comment on the validity of perturbative expansion.
If we start at time $\tau = \tau_1$ such that $A(\tau_1)$ is of order one, then moving forward in time
the perturbations ${\widetilde h}_{MN}$ decay compared with the metric they are expanded around, so the perturbative
expansion remains valid.

\section{Relation to 't Hooft's Work}

{}Thus, as we saw in the previous section, spontaneous breaking of spatial diffeomorphisms indeed results
in massive gravity in an expanding background. In \cite{thooft} a somewhat different setup was discussed in
the context of obtaining massive gravity. Here we would like to review 't Hooft's work in the language of this
paper to make a connection.

{}Thus, consider the following action:
\begin{equation}
 S_1=M_P^{D-2}\int d^Dx \sqrt{-G}\left[ R - Z_{AB} \nabla^M \phi^A \nabla_M \phi^B - \Lambda\right]~,
 \label{actionphi1}
\end{equation}
where $\Lambda$ is the cosmological constant, and $Z_{AB}$ is a constant metric for the scalar sector, where
$A = 0,\dots,D-1$. In the following we will take this metric to be identical to the Minkowski metric: $Z_{AB} =
{\delta_A}^M{\delta_B}^N \eta_{MN}$.

{}The equations of motion read:
\begin{eqnarray}
 \label{phi11}
 && \nabla^2 \phi^A = 0~,\\
 \label{einstein11}
 && R_{MN} - {1\over 2}G_{MN} R = \nonumber\\
 &&Z_{AB} \left[\nabla_M\phi^A \nabla_N\phi^B
 -{1\over 2}G_{MN}Z_{AB} \nabla^S \phi^A \nabla_S \phi^B \right] -{1\over 2}G_{MN}\Lambda~.
\end{eqnarray}
There is a solution to this system with flat Minkowski metric:
\begin{eqnarray}
 &&\phi^A = m~{\delta^A}_M~x^M~,\\
 &&G_{MN} = \eta_{MN}~,
\end{eqnarray}
where
\begin{equation}
 m^2 = -\Lambda / (D-2)~.
\end{equation}
The scalar fluctuations $\varphi^A$ can be gauged away using the diffeomorphisms:
\begin{equation}\label{diffphi1}
 \delta\varphi^A =\nabla_M \phi^A \xi^M = m~{\delta^A}_M ~\xi^M~.
\end{equation}
The linearized equations of motion (\ref{phi11}) and (\ref{einstein11}) then read
($G_{MN} = \eta_{MN} + h_{MN}$):
\begin{eqnarray}
 &&2\partial^N h_{MN} - \partial_M h = 0~,\\
 &&\partial_S\partial^S h_{MN} +\partial_M\partial_N
 h-\partial_M \partial^S h_{SN}-
 \partial_N \partial^S h_{SM}-\eta_{MN}
 \left[\partial_S\partial^S h-\partial^S\partial^R
 h_{SR}\right] = \nonumber\\
 &&m^2\left[2h_{MN} - \eta_{MN} h\right]~,
\end{eqnarray}
where $h\equiv h^M_M$.

{}We can now see the unitarity issue discussed in \cite{thooft}. To do this, let us consider a more
general set of equations:
\begin{eqnarray}\label{condh}
 &&\zeta\partial^N h_{MN} - \partial_M h = 0~,\\
 &&\partial_S\partial^S h_{MN} +\partial_M\partial_N
 h-\partial_M \partial^S h_{SN}-
 \partial_N \partial^S h_{SM}-\eta_{MN}
 \left[\partial_S\partial^S h-\partial^S\partial^R
 h_{SR}\right] = \nonumber\\
 &&m^2\left[\zeta h_{MN} - \eta_{MN} h\right]~,
\end{eqnarray}
where $\zeta$ is a parameter. Taking the trace of the second equation, we have:
\begin{equation}
 (D-2)\left[\partial^S\partial^R h_{SR} -  \partial^S\partial_S h\right] = -m^2 (D-\zeta)h~.
\end{equation}
On the other hand,
\begin{equation}
 \zeta \partial^M\partial^N h_{MN} = \partial^M\partial_M h~,
\end{equation}
so we have the following equation of motion for $h$:
\begin{equation}
 (D-2)(1-1/\zeta)\partial^S\partial_S h = m^2 (D-\zeta) h~.
\end{equation}
This means that, unless $\zeta = 1$, $h$ is a propagating degree of freedom, and since this degree of
freedom has negative norm, the corresponding theory is non-unitary. The number of degrees of freedom in
this model is $D(D+1)/2$ (from $h_{MN}$) less $D$ (from the condition (\ref{condh})), which gives $D(D-1)/2$.
This is massive gravity plus an {\em undecoupled} trace component $h$, a non-unitary theory. At the special
value of $\zeta =1$ we have $h = 0$, and the number of propagating degrees of freedom is
$D(D-1)/2 - 1 = (D+1)(D-2)/2$, which is the number of degrees of freedom of a massive graviton.

{}Another way to view
the reason for non-unitarity in this model is to note that we have
indefinite metric $Z_{AB}$ for the scalar sector\footnote{Actually, in \cite{thooft} the metric $Z_{AB}$
is positive definite, but one of the scalars has imaginary VEV, $\phi^0 = imt$, with the same net effect.}.
Indeed, effectively, we just have a massless {\em vector}
meson $\phi^M$ with the Lagrangian $L_1\sim - \nabla^M \phi^N \nabla_M \phi_N$. This Lagrangian is clearly
non-unitary, hence the issue. In \cite{thooft} two ways of removing this non-unitarity were discussed.
One is to ensure that the matter energy-momentum tensor does not couple to the
non-unitary mode $h$ by assuming that the corresponding coupling has a special form at the classical level. This
coupling is expected to be modified at the quantum level. Another possibility discussed in \cite{thooft} is to
remove the time-like component $\phi^0$ via a non-linear constraint.

{}Here we would like to mention yet another possibility, which is a well-known
approach for making a vector meson theory unitary -- by turning it into a gauge field. We then have the Lagrangian
$L_2 \sim -F^{MN}F_{MN}$ with $F_{MN} = \nabla_M A_N - \nabla_N A_M$. However, with
the massless gauge field we will be able to break diffeomorphisms in $(D-2)$ {\em spatial} directions by finding
solutions with constant $F_{MN}$. (For instance, in $D=4$, we have two independent vectors, the constant electric field
${\vec E}$ and the constant magnetic field ${\vec B}$, and these are related to two spatial directions in which
diffeomorphisms are broken.)In any case, the key here is that unitarity is broken as soon as we break the time-like
diffeomorphism. In the previous section we avoided this altogether by breaking only
$(D-1)$ diffeomorphisms in the spatial directions, and we obtained massive gravity
with just the correct count of the propagating degrees of freedom.



\newpage
\begin{thebibliography}{99}

\bibitem{KL} Z. Kakushadze and P. Langfelder,
``Gravitational Higgs Mechanism", Mod. Phys. Lett. A15 (2000) 2265, arXiv:hep-th/0011245.

\bibitem{thooft} G. 't Hooft,
``Unitarity in the Brout-Englert-Higgs Mechanism for
Gravity", arXiv:0708.3184 [hep-th].

\bibitem{RS} L. Randall and R. Sundrum,
``An Alternative to Compactification", Phys. Rev. Lett. 83 (1999) 4690, arXiv:hep-th/9906064.

\bibitem{Por} M. Porrati,
``Higgs Phenomenon for 4-D Gravity in Anti de Sitter Space", JHEP 0204 (2002) 058, arXiv:hep-th/0112166.

\bibitem{Ch} A.H. Chamseddine,
``Spontaneous Symmetry Breaking for Massive Spin-2 Interacting with Gravity", Phys. Lett. B557 (2003) 247, arXiv:hep-th/0301014.

\bibitem{Kir} I. Kirsch,
``A Higgs Mechanism for Gravity", Phys. Rev. D72 (2005) 024001, arXiv:hep-th/0503024.

\bibitem{GT} M.B. Green and C.B. Thorn,
``Continuing between Closed and Open Strings", Nucl. Phys. B367 (1991) 462.

\bibitem{Siegel} W. Siegel,
``Hidden Gravity in Open-String Field Theory", Phys. Rev. D49 (1994) 4144, arXiv:hep-th/9312117.

\bibitem{Gab} G. Gabadadze,
``Cargese Lectures on Brane Induced Gravity", arXiv:0705.1929 [hep-th].

\bibitem{Jackiw} R. Jackiw,
``Lorentz Violation in Diffeomorphism Invariant Theory", arXiv:0709.2348 [hep-th].

\bibitem{Acc} A.G. Riess {\em et al}.,
``Observational Evidence from Supernovae for an Accelerating Universe and a Cosmological Constant",
Astron. J. 116 (1998) 1009, arXiv:astro-ph/9805201;\\
S. Perlmutter {\em et al}.,
``Measurements of Omega and Lambda from 42 High-Redshift Supernovae ",
Astrophys. J. 517 (1999) 565, arXiv:astro-ph/9812133.

\end{thebibliography}
\end{document}